\documentclass[11pt]{cernrep} \usepackage{graphicx}

\begin{document}

\title{Quark antiquark energies and the screening mass in a Quark-Gluon plasma
  at low and high temperatures}

\author{Olaf Kaczmarek$^{1}$ and Felix Zantow$^{2}$}

\institute{
  $^1$ Fakult\"at f\"ur Physik, Universit\"at Bielefeld, D-33615 Bielefeld, Germany \\
  $^2$ Physics Department, Brookhaven Natl. Laboratory, Upton, NY-11973, USA }

%%%%%%%%%%%%%%%%%%%%%%%%%%%%%%%%%%%%%%%%%%%%%%%%%%%%%%%%%%%%%%
% Only one use of \author and \institute is recognised - you 
% can do multiple affiliations eg as
%
% \author{T. DeYoung$^{1}$ and G. C. Hill$^{2}$ for the AMANDA Collaboration}
% \institute{
% $^1$ Santa Cruz Institute for Particle Physics, University of California,
% Santa Cruz, CA 95064, USA \\
% $^2$ Department of Physics, University of Wisconsin, Madison, WI 53706, USA
%  }
%%%%%%%%%%%%%%%%%%%%%%%%%%%%%%%%%%%%%%%%%%%%%%%%%%%%%%%%%%%%%%

\maketitle

\begin{abstract}
  We discuss quark antiquark energies and the screening mass in hot QCD using
  the non-perturbative lattice approach. For this purpose we analyze properties
  of quark antiquark energies and entropies at infinitely large separation of
  the quark antiquark pair at low and high temperatures. In the limit of high
  temperatures these energies and entropies can be related perturbatively to
  the temperature dependence of the Debye mass and the coupling. On the one
  hand our analysis thus suggests that the quark antiquark energies at
  (infinite) large distances are rather related to the Debye screening mass and
  the coupling than to the temperature dependence of heavy-light meson masses.
  On the other hand we find no or only little differences in all mass scales
  introduced by us when changing from quenched to 2-flavor QCD at temperatures
  which are only moderately above the phase transition.
\end{abstract}

\section{The Debye screening mass}
Color screening has been considered as an important mechanism for
deconfinement, and in this context, quarkonium suppression as signal for
quark-gluon plasma production in heavy ion collision experiments
\cite{Satz,Karsch}. The temperature dependence of the Debye screening mass,
$m_D(T)$, provides an important measure for the strength of the screening
property of the QCD medium. Due to difficulties when calculating the Debye mass
in perturbation theory at temperatures which are only moderately above the
phase transition temperature ($T_c$), the calculation of the screening mass has
been to large extent considered using the non-perturbative lattice approach. In
this approach one usually discusses either the infrared limit of the gluon
propagator and/or the large distance color screened Coulomb behavior of quark
antiquark free energies
\cite{Heller:1997nq,Kaczmarek:1999mm,Digal:2003jc,Nakamura:2004wr,Kaczmarek:2004gv,Kaczmarek:2005ui}.
We will discuss here the property that the screening mass can also be related
at asymptotic high temperatures to the large distance behavior of quark
antiquark energies and entropies
\cite{Kaczmarek:2002mc,Kaczmarek:2005gi,Kaczeetal,KaczeEPJC1}. In particular,
we conclude that the finite values of the finite temperature energies which are
approached at (infinitely) large distances are rather related to the
temperature dependence of the Debye mass and the coupling than to the
temperature dependence of masses of corresponding heavy-light meson systems.

\subsection{$Q\bar Q$ Free Energy, Entropy and the Screening mass}
The presence of static color charges will polarize the medium and the parton
density will change compared to the density distribution of the heat bath
without containing static charges. In particular, the parton density is
considered to rapidly increase in the close vicinity of the static charges and
the energy and entropy which is needed to neutralize the additional charges
will in general depend on temperature ($T$) and the distance ($r$) between the
static charges. This property is illustrated in Fig.~1 (see also our discussion
in Ref.~\cite{KaczePoS}): In the limit of large distances ($r\to\infty$) the
parton clouds can be considered as well separated (a). In this case the change
in energy (free or internal energy) and entropy due to the presence of static
charges in the heat bath will only depend on temperature\footnote{Of course
  this property changes qualitatively in quenched QCD at temperatures below
  deconfinement where the confinement forces should be related to a {\em
    string} (which cannot break in quenched QCD) rather than to {\em clouds}
  (which can be separated).}. While the change in free energy at (asymptotic)
large distances, $F_\infty(T)\equiv \lim_{r\to\infty}F_1(r,T)$, is a steadily
decreasing function with increasing temperatures, the corresponding change in
internal energy, $U_\infty(T)\equiv-T^2\partial (F_\infty/T)/\partial T$, and
entropy, $S_\infty(T)\equiv-\partial F_\infty/\partial T$, behave more
complicated as function of temperature \cite{Kaczmarek:2005gi,KaczePoS2}. In
particular, the energy and entropy which are needed to neutralize the static
charges in the heat bath show a strong increase at temperatures close to the
transition temperature and exhibit a sharp peak at the phase transition
\cite{Kaczmarek:2005gi,KaczePoS2}. At high temperatures, however, the internal
energy is continuously decreasing with increasing temperatures while the
entropy vanishes slowly in the limit of high temperatures. This behavior is
also expected at asymptotic high temperatures from perturbation theory
(following \cite{Kaczmarek:2005gi,Gava}) where the temperature dependence of
$F_\infty(T)$, $U_\infty(T)$ and $S_\infty(T)$ can be related to the Debye mass
($m_D$) and the coupling ($\alpha=g^2/(4\pi)$), {\em i.e.}
\begin{eqnarray}
F_\infty(T)&\simeq&-\frac{4}{3}m_D(T)\alpha(T)\label{finf}\;,\\
U_\infty(T)&\simeq&4m_D(T)\alpha(T)\frac{\beta(g)}{g(T)}\;\simeq\;-{\cal O}(Tg^5)\label{uinf}\;,\\
S_\infty(T)&\simeq&+\frac{4}{3}\frac{m_D(T)}{T}\alpha(T)\label{sinf}\;,
\end{eqnarray}
where $\beta(g)$ denotes the QCD $\beta$-function. These relations already
indicate that it could be misleading to relate the large distance values of the
finite temperature energies to the temperature dependence of masses of
corresponding heavy-light meson systems.

When going to smaller distances ($r<\infty$), however, the parton clouds will
overlap (see Fig.~1: b, c) and the finite temperature energies and the entropy
will also depend on distance, {\em i.e.} $F_1\equiv F_1(r,T)$, $U_1\equiv
U_1(r,T)$ and $S_1\equiv S_1(r,T)$. In this case high temperature perturbation
theory suggests a color screened Coulombic behavior for the $r$-dependence of
the free energy, {\em i.e.}
\begin{eqnarray}
\delta F_1(r,T)&\equiv&F_1(r,T)-F_\infty(T)\;\simeq\;-\frac{4}{3}\frac{\alpha(T)}{r}e^{-m_D(T)r}\;.\label{fit}
\end{eqnarray}
In the limit of small distances ($r\to0$), however, the quark antiquark free
energy is given by the heavy quark potential, $V(r)$, at zero temperature,
$F_1(r\ll1/T,T)\simeq V(r)$. This property can be used to fix the finite
temperature energies at small distances to the heavy quark potential. Once the
free energy is fixed at small distances also the large distance behavior is
properly determined, {\em i.e.} $F_\infty(T)$ is properly fixed, and it can no
longer be assumed that the finite temperature energies and the entropy approach
zero at large distances \cite{Kaczmarek:2002mc}.

\begin{figure}[t]
\begin{center}
  \includegraphics[width=4.in]{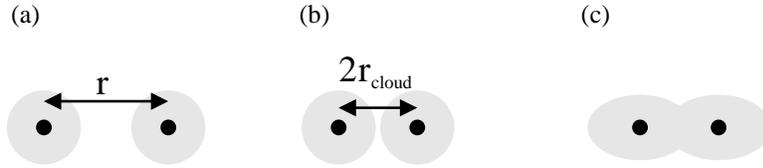}
\end{center}
\vskip -0.3cm
\caption{Illustration of screening through the polarization of gluons in the
  medium (taken from Ref.~\cite{KaczePoS}): (a) At large distances
  ($r\to\infty$) the gluon clouds which surround the static test charges are
  well separated. At smaller distances the clouds begin to overlap (b) and the
  geometric structure of the clouds will also depend on the separation between
  the test charges (c).}
\vskip -0.3cm
\end{figure}

\subsection{Lattice Results}
We have calculated quark antiquark free energies as well as the corresponding
internal energies and entropies at several temperatures below and above the
deconfinement phase transition using the non-perturbative lattice approach in
quenched \cite{Kaczmarek:2004gv,Kaczmarek:2002mc,Kaczeetal} and $2$-flavor QCD
\cite{Kaczmarek:2005ui,Kaczmarek:2005gi,Kaczeetal2}. Above $T_c$, we extracted
the Debye mass, $m_D(T)$, using a best fit analysis of the color singlet free
energy with a color screened Coulomb form at large distances with respect to
Eq.~(\ref{fit}). Here, we may also define certain mass scales,
$\mu_{F_\infty}(T)$ and $\mu_{S_\infty}(T)$, by using Eqs.~(\ref{finf},
\ref{sinf}), {\em i.e.} we use the temperature dependence of the large distance
plateau values of the free energy ($F_\infty$) and entropy ($S_\infty$) and
define
\begin{eqnarray}
\mu_{F_\infty}(T)\;\equiv\;-\frac{3}{4\alpha(T)}\times F_\infty(T)\;,\quad
\mu_{S_\infty}(T)\;\equiv\;+\frac{3}{4\alpha(T)}\times
TS_\infty(T)\;.\label{scales}
\end{eqnarray}
In the limit of high temperatures both, $\mu_{F_\infty}(T)$ and
$\mu_{S_\infty}(T)$, are expected to characterize the temperature dependence of
the Debye mass, $m_D(T)$.

Of course, both mass scales defined in (\ref{scales}) will depend on the proper
definition of the coupling which enters (\ref{scales}) and is used to fix
$\mu_{F_\infty}(T)$ and $\mu_{S_\infty}(T)$. Assuming vanishing quark masses in
QCD, we fix $\mu_{F_\infty}(T)$ and $\mu_{S_\infty}(T)$ using the temperature
dependence of the $2$-loop running coupling in the $\overline{MS}$-scheme,
\begin{eqnarray}
g_{2-loop}^{-2}(T)&=&2\beta_0\ln\left(\frac{\mu
    T}{\Lambda_{\overline{MS}}}\right)+\frac{\beta_1}{\beta_0}\ln\left(2\ln\left(\frac{\mu
    T}{\Lambda_{\overline{MS}}}\right)\right)\;,\label{2loop}
\end{eqnarray}
with
\begin{eqnarray}
\beta_0\;=\;\frac{1}{16\pi^2}\left(11-\frac{2N_f}{3}\right)\;,\quad\;\beta_1\;
=\;\frac{1}{(16\pi^2)^2}\left(102-\frac{38N_f}{3}\right)\;,
\end{eqnarray}
where we used $T_c/\Lambda_{\overline{MS}}=1.14(4)$ \cite{z8,z21} in quenched
($N_f=0$) and $T_c/\Lambda_{\overline{MS}}=0.77(21)$ \cite{z22,z53} in
$2$-flavor ($N_f=2$) QCD. In both cases, quenched and $2$-flavor QCD, we fixed
the renormalization scale $\mu$ to be $2\pi$. Despite the dependence of the
mass scales introduced in (\ref{scales}) from the definition of the coupling,
$\mu_{F_\infty}(T)$ will also depend on the fixing of the heavy quark potential
at zero temperature, $V(r)$, which is used for renormalization of the free
energy and could add an overall constant contribution to $F_\infty(T)$. Here
and in what follows we fixed the heavy quark potential as described in
Refs.~\cite{Kaczmarek:2002mc,KaczeEPJC2}. Our results for the different mass
scales $\mu_{F_\infty}(T)$ and $\mu_{S_\infty}(T)$ are shown in Fig.~2 as
function of temperature, $T/T_c$, for quenched (filled symbols) and $2$-flavor
QCD (open symbols) at temperatures ranging from $T_c$ up to temperatures about
$5.5T_c$. We also compare in that figure $\mu_{F_\infty}(T)$ and
$\mu_{S_\infty}(T)$ to the temperature dependence of the Debye mass, $m_D(T)$,
recently obtained in quenched \cite{Kaczmarek:2004gv} and $2$-flavor QCD
\cite{Kaczmarek:2005ui}.

Following the discussions of the Debye mass in
Refs.~\cite{Kaczmarek:2004gv,Kaczmarek:2005ui}, $m_D(T)$ is about 500 MeV at
temperatures close above the phase transition and continuously increases with
increasing temperatures to about 2000 MeV at $4T_c$. In fact, the steadily
increasing Debye mass can be well described with a perturbative inspired
ansatz,
\begin{eqnarray}
m_D(T)&=&A_{N_f}\sqrt{1+\frac{N_f}{6}}\; g(T)\;T\;,\label{PTI}
\end{eqnarray}
allowing for a non-perturbative overall multiplicative correction, $A_{N_f}$,
and using for the temperature dependence of the coupling, $g(T)$, the
perturbative $2$-loop running coupling given in Eq.~(\ref{2loop}) and a
renormalization scale $\mu=2\pi$. For temperatures moderately above the
transition one finds $A_{N_f=0}\simeq1.51(2)$ in quenched
\cite{Kaczmarek:2004gv} and $A_{N_f=2}\simeq1.42(2)$ in $2$-flavor QCD
\cite{Kaczmarek:2005ui}. The latter estimate for $m_D(T)$ in $2$-flavor QCD is
shown in Fig.~2 as solid lines (including the error on $A_{N_f=2}$). We again
stress here (see also our discussion of the screening mass and length in
Refs.~\cite{Kaczmarek:2005ui,KaczePoS,KaczePoS2}) we see no or only little
differences between $m_D(T)$ calculated in quenched and $2$-flavor QCD in the
entire temperature range shown in Fig.~2. A different discussion concerning
this issue has recently been given in Refs.~\cite{P1,P2}.

\begin{figure}[t]
\begin{center}
  \includegraphics[width=4.in]{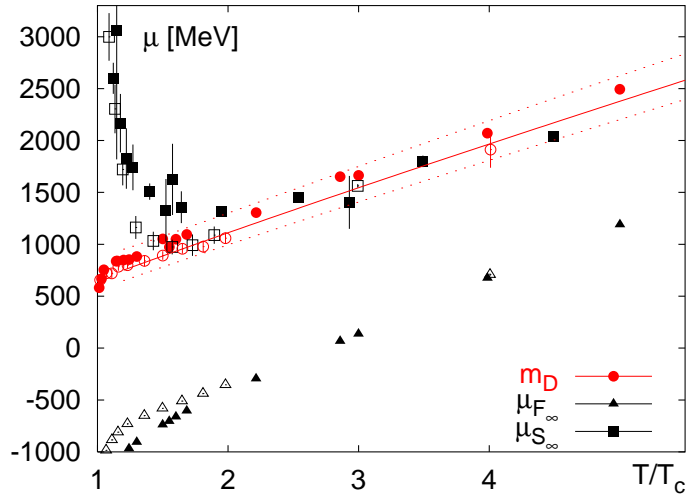}
\end{center}
\vskip -0.3cm
\caption{The Debye screening mass, $m_D(T)$, obtained in quenched (filled
  circles) \cite{Kaczmarek:2004gv} and $2$-flavor QCD (open circles)
  \cite{Kaczmarek:2005ui} as function of $T/T_c$. To convert $m_D(T)$ to
  physical units we used $T_c=270$ MeV for quenched and $T_c=202$ MeV for
  $2$-flavor QCD. The lines indicate a best-fit analysis of the perturbatively
  inspired ansatz (\ref{PTI}) for $m_D(T)$ in $2$-flavor QCD (see
  Ref.~\cite{Kaczmarek:2005ui} for further details). We also show here lattice
  results for the different mass scales $\mu_{F_\infty}(T)$ and
  $\mu_{S_\infty}(T)$ introduced in the text (see Sec.~1) obtained in quenched
  (filled symbols) and $2$-flavor QCD (open symbols). }
\vskip -0.3cm
\end{figure}
We also show in Fig.~2 the temperature dependence of the mass scales
$\mu_{F_\infty}(T)$ and $\mu_{S_\infty}(T)$ obtained from the infinite large
distance values of the quark antiquark free energy and entropy. At low
temperatures, {\em i.e.} temperatures below $3T_c$, $\mu_{F_\infty}(T)$
indicates negative values while it changes sign at about $3T_c$ and
continuously increases with increasing temperature. In fact, below $3T_c$ the
value which is approached by the free energy at large distances, $F_\infty(T)$,
is still positive and thus may indicate a temperature regime which is not quite
accessible with leading order perturbation theory, {\em i.e.} the leading order
approximation given in Eq.~(\ref{finf}) does indicate a misleading sign for
$F_\infty(T)$ in this temperature regime (below $3T_c$). At low temperatures it
thus makes no sense to identify $\mu_{F_\infty}(T)$ with a scale that
characterizes the Debye mass. The mass scale, however, which is obtained from
the finite values of the quark antiquark entropy, $\mu_{S_\infty}(T)$, is also
shown in Fig.~2 for quenched (filled squares) and full QCD (open squares) and
does not depend on the fixing of $V(r)$. At temperatures close above the
deconfinement phase transition $\mu_{S_\infty}(T)$ approaches unexpected values
about $3000$ MeV. It, however, rapidly drops below $1500$ MeV already at
temperatures about twice as large than $T_c$ and than starts to increase with
increasing temperatures as expected in perturbation theory for a mass scale
that might mimic the temperature dependence of the Debye mass. In particular,
at temperatures larger than $2T_c$ the mass scales $\mu_{S_\infty}(T)$ and
$m_D(T)$ become of similar magnitude in the entire temperature range analyzed
here. Again, when comparing our results for $\mu_{F_\infty}(T)$ and
$\mu_{S_\infty}(T)$ obtained in quenched and $2$-flavor QCD we find no or only
little differences at temperatures which are only moderately above the
transition.

\section{Summary}
The present status of our analysis of the quark antiquark free energy, internal
energy and the entropy at finite temperature concerning Debye screening effects
suggest the following two statements (see also
Refs.~\cite{Kaczmarek:2005ui,Kaczmarek:2005gi,KaczePoS}): (i) It might be
misleading to relate the large distance values of the finite temperature
energies to the mass dependence of the corresponding heavy-light meson systems;
in particular, the temperature dependence of $F_\infty(T)$ and $U_\infty(T)$
might be rather related to the temperature dependence of the Debye mass and the
coupling. (ii) The large differences between Debye screening effects observed
in Refs.~\cite{P1,P2} when changing from quenched to $3$-flavor QCD are not
apparent in any of the scales analyzed here when changing from quenched to
$2$-flavor QCD at temperatures which are only moderately above the transition
temperature.

\section*{Acknowledgments}
We thank F. Karsch, E. Laermann and H. Satz for helpful discussions. This work
has partly been supported by DFG under grant FOR 339/2-1 and by BMBF under
Grant No.06BI102 and partly by Contract No. DE-AC02-98CH10886 with the U.S.
Department of Energy.

\end{document}